# Shell Structure and Strengthening of Superconducting Pair Correlation in Nanoclusters

Vladimir Z. Kresin[a] and Yurii N. Ovchinnikov[b,c],

[a] Lawrence Berkeley Laboratory, University of California, Berkeley CA 94720

[b] L.D.Landau Institute for Theoretical Physics, Russian Academy of Sciences, 117334 Moscow, Russia

[c] Max-Planck Institute for the Physics of Complex Systems, Dresden, D-01187, Germany

## Abstract

The existence of shell structure and the accompanying high degeneracy of electronic levels leads to the possibility of strong superconducting pairing in metallic nanoclusters with $N \sim 10^2$-$10^3$ delocalized electrons. The most favorable cases correspond to (a) "magic" clusters with strongly degenerate highest occupied and lowest unoccupied shells and a relatively small energy spacing between them as well as to (b) clusters with slightly incomplete shells and small Jahn-Teller splitting. It is shown that realistic sets of parameters lead to very high values of $T_c$ as well as to a strong alteration of the energy spectrum. The impact of fluctuations is analyzed. Spectroscopic experiments aimed at detecting the presence of pair correlations are proposed. The pairing should also manifest itself via odd-even effects in cluster spectra, similar to the case in nuclei.

I. **Introduction**

The problem of the superconducting state of metallic nanoparticles has attracted a lot of interest (see, e.g., [1-12] and the reviews [13,14]). However, the experimental and theoretical studies have focused, mainly, on relatively large nanoparticles ($N \gtrsim 10^4 - 10^5$, $N$ is the number of delocalized electrons). In this paper we consider superconducting pairing in small nanoclusters with $N \approx 10^2 - 10^3$. Our approach was previously outlined in short communications [15]. Here we present a detailed description and some new results.

The most distinguished feature of small nanoparticles is the discrete nature of the electronic spectrum. In dealing with small nanoclusters, $N \approx 10^2 - 10^3$, one might think [13] that they do not display superconducting properties, because the average level spacing ($E_F/N \sim 10^2 - 10^3$ K) greatly exceeds the pairing energy gap. However, the presence of the so-called shell structure leads to a more complex situation. It turns out that for many real clusters the pattern of electronic states is very different from an equally spaced level distribution. Instead, they contain highly degenerate energy levels, or groups of very close levels, so that the energy spacing for electronic states close to the Fermi level, $E_F$, is rather small. The situation in such clusters is very favorable for pairing and one can even expect, as will be shown below, a giant strengthening of the phenomenon relative to bulk samples.

The structure of the paper is as follows. Sec. II contains qualitative description of the pairing in nanoclusters and the impact of shell structure. The main equations are introduced in Sec.III. A detailed evaluation of the critical temperature and the energy spectrum in the presence of pairing is given in Secs. IV and V. Sec V also contains an analysis of the properties near $T_C$ and a discussion of the role of fluctuations. Pair correlation in some specific clusters is described in Sec.VI. The manifestations of pair correlation in small nanoclusters and the possibility of their experimental observation are discussed in Sec. VII.

II. **Shell structure and pairing: Qualitative picture**

Metallic clusters contain delocalized electrons, and their states form shells similar to those in atoms or nuclei. This shell structure has been extensively studied in both experiment and theory (see, e.g., the reviews [16-18]). In addition to the alkali clusters in which shell structure was originally discovered [19], its presence has been detected also for many other nanoclusters, such as Al, Ga, Zn, Cd, In, etc. [20-25]. Shell structure has been observed in clusters containing up to hundreds of delocalized electrons, and recently reported even in larger Ga clusters, $N \cong 7 \cdot 10^3$ [22].

The presence of shell structure is manifested in the appearance of so-called "magic" numbers with $N=N_m$, e.g., $N_m$ = 20, 40, 58, …, 138, 168, …, etc. Such clusters possess completely filled electronic shells and, similarly to "inert" atoms, are most stable. The "magic" clusters have spherical shapes and their electronic states are labeled by their orbital momentum (*l*) and radial quantum number (*n*).

If the shell is not compete, the cluster undergoes a Jahn-Teller distortion and its shape becomes ellipsoidal. This splits the degenerate levels into sub-levels labeled by the projection of the orbital momentum (*m*). For our purposes, the most interesting case corresponds to clusters containing a slightly incomplete highest occupied shell (HOS). Then one can expect a small level splitting ( see Fig. 1). Note that in addition to the Jahn-Teller distortion, an additional splitting might occur because of the exchange interaction [28]. However, we do not include spin-orbit effects in our treatment below.

Thus the presence of shell structure does , indeed, make the distribution of energy levels far from equidistant.

The importance of shell structure and the corresponding degeneracy was discussed in [ 6,7,11,12 ]. In an especially interesting paper [7], motivated by the discovery of $C_{60}$ clusters and fulleride superconductivity in fullerines , it was proposed that spherical clusters with a half-filled shells should possess high values of $T_c$. However, in this situation the cluster shape deformation would be very large, drastically decreasing $T_c$. The author [7] suggested that it might be possible to employ a cluster network incorporating charge transfer to overcome this problem. The idea of such superconducting molecular (or, more exactly, cluster) crystal is very interesting. In the present paper we concentrate on pairing in an isolated cluster. We will consider clusters with almost filled shells to avoid large deformation.

The superconducting pairs are formed by electrons with opposite projection of angular momentum ($m$, $-m$). In many aspects the picture of pairing is similar to that in atomic nuclei (see, e.g., [29] and the review [30]). In both cases (nuclei and clusters) we are dealing with finite Fermi systems and shells structures. The pairing states are labeled by similar quantum numbers ($m$, $-m$). The manifestation of pairing also has similarities (see below, Sec. VII). However, for clusters we can develop a more microscopic approach, thanks to action of Coulomb forces and presence of two subsystems (electrons and ions). The pairing is caused by the electron-vibrational coupling, i.e., the mechanism is similar to that in usual bulk superconductors.

The most favorable case corresponds to clusters in which the HOS and LUS have high degeneracies and, in addition, the spacing between them is relatively small. The pairing effect strongly influences the cluster's energy spectrum, with the impact particularly strong for clusters with slightly incomplete shells (Fig. 1) where the excitation energy in the absence of pairing can be rather small. A detailed theory will be described in the next section.

One should stress that the strength of pair correlation varies for different clusters. Correspondingly, the critical temperature, $T_C$, and the energy gap are strongly dependent upon the cluster's parameters, its shape, the strength of the coupling, etc. We will consider a number of specific cases in the subsequent sections. It is important that in some special, but perfectly realistic situations one can obtain a very large values of $T_C$. Qualitatively, this can be understood in the following way. If the HOS is highly degenerate, this means that this shell contains many electrons, which can be viewed as a sharp peak in the density of states at the Fermi level. Situation is similar to that studied in [31] for bulk materials; the presence of a peak in the density of states results in a noticeable increase in $T_C$.

One may also recall that generally speaking, it is known, size quantization leads to an effective increase in density of states, which can lead to an increase in $T_c$. (see, e.g., [32,33]). Shell structure in clusters corresponds to an extreme case of size quantization, when the usual size effect is enhanced by the large degeneracy of electronic states. If, in addition, the energy spacing is relatively small (see below), this leads to a large enhancement of the pairing phenomenon.

III. **Main equations.**

Let us write down a general equation describing the pairing in a metallic cluster. The electron-vibrational interaction, like in the usual case, is considered as the major mechanism of pairing. However, as is known, the BCS formalism is valid in weak coupling approximation (then $T_c \ll \tilde{\Omega}$, where $\tilde{\Omega}$ is the characteristic vibrational frequency). Since we want to go beyond this restriction, we start with a more general equation (cf.[34-36]) which explicitly contains the vibrational propagator. The equation for the pairing order parameter $\Delta(\omega_n)$ has the following form:

$$\Delta(\omega_n)Z = \eta \frac{T}{2V} \sum_{\omega_{n'}} \sum_s D(\omega_n - \omega_{n'}) F_s^+(\omega_{n'}) \qquad (1)$$

Here $\omega_n = (2n+1)\pi T$; n = 0, ± 1, ± 2,… (we employ the thermodynamic Green's functions formalism, see, e.g. [37]),

$$D(\omega_n - \omega_{n'}, \tilde{\Omega}) = \tilde{\Omega}^2 \left[ (\omega_n - \omega_{n'})^2 + \tilde{\Omega}^2 \right]^{-1} \qquad (1')$$

is the vibrational propagator, and

$$F_s^+(\omega_{n'}) = \Delta(\omega_{n'}) \left[ \omega_{n'}^2 + \xi_s^2 + \Delta^2(\omega_{n'}) \right]^{-1} \qquad (1'')$$

is the Gor'kov pairing function [38], $\xi_s = E_s - \mu$ is the energy of the s'th electronic state referred to the chemical potential μ, V is the cluster volume, $\eta = <I>^2 / M\tilde{\Omega}^2$ is the Hopfield parameter, <I> is the electron-ion matrix element averaged over electronic states involved in the pairing (see, e.g. [39,40]), M is the ionic mass, and Z is the renormalization function which describes the electron-vibrational scattering and is given by

$$Z = 1 + \eta \frac{T}{2V\omega_n} \sum_{\omega_n} \sum_{s} D(\omega_n - \omega_{n'}, \tilde{\Omega}) \omega_{n'} (\omega_{n'}^2 + \xi_s^2 + \Delta^2(\omega_{n'}))^{-1} \qquad (2)$$

Eqs. (1), (2) contain a summation over all discrete electronic states. For "magic" clusters which have a spherical shape, one can replace summation over states by summation over the shells: $\sum_s \to \sum_j G_j$, where $G_j$ is the shell degeneracy: $G_j=2(2l_j+1)$, $l_j$ is the orbital momentum. If the shell is incomplete, the cluster undergoes a Jahn-Teller deformation, so that its shape becomes ellipsoidal, and the states "s" are classified by their projection of the orbital momentum $|m| \leq l$, so that each level contain up to four electrons (for $|m| \geq 1$ ). Note that in a weak coupling case ($\eta/V \ll 1$ and correspondingly $\pi T_C \ll \tilde{\Omega}$), one should put in Eqs. (1,2) $Z=1$, $D=1$, recovering the usual BCS scheme.

Equation (1) looks similar to that in the theory of strong coupling superconductivity [41], but is different in two key aspects. Firstly, it contains a summation over discrete energy levels $E_S$ whereas for a bulk superconductor one integrates over continuous energy spectrum (over $\xi$). Another important difference is that, as opposed to a bulk superconductor, here we are dealing with a finite Fermi system, so that a number of electrons N is fixed. As a result, the position of the chemical potential differs from the Fermi level $E_F$ and is determined by the values of N and T. Specifically, one can write

$$N = \sum_{\omega_n} \sum_{s} \Im_s(\omega_n) e^{i\omega_n \tau}|_{\tau \to 0} \qquad (3)$$

Here $\Im_S(\omega_n)$ is the thermodynamic Green's function: $J_s = (i\omega_n + \xi)(\omega_n^2 + \xi_s^2 + \Delta^2(\omega_n))^{-1}$.

Performing the summation, we obtain

$$N = \sum_s \left(u_s^2 \varphi_s^- + v_s^2 \varphi_s^+\right) \tag{4}$$

where

$$u_s^2, v_s^2 = 0.5(1 \mp \xi_s / \varepsilon_s) \quad ; \quad \varphi_s^{\mp} = \left[1 + \exp(\mp \varepsilon_s / T)\right]^{-1} \tag{4'}$$

and

$$\varepsilon_s = \left(\xi_s^2 + \varepsilon_{0;s}^2\right)^{1/2} \tag{4''}$$

$\varepsilon_{0;s}$ is the gap parameter for the s'th level and $\varepsilon_{0;s}$ is the root of the equation: $\varepsilon_{0;s} = \Delta(i\varepsilon_s)$. Since $\xi_s = E_s - \mu$, Eq. (4) determines the position of the chemical potential for the given number of electrons N as well as the dependence $\mu(T)$.

Note also that for clusters of interest ($N \gtrsim 10^2$; then $k_H R >> 1$) the order parameter $\Delta \equiv \Delta(\omega_n)$; the coordinate dependence and consequently the dependence on $s$ is rather weak (see, e.g.,[10,42]) and can be neglected. Here R is the cluster radius, and $k_H$ is the electronic wave vector for the highest occupied shell ($k_H \cong 2/r_s$, where $r_s$ is the electron density parameter; we put $\hbar = 1$). The value of $k_H$ for the clusters of interest is close to the Thomas-Fermi screening wave vector and to the Fermi momentum $k_F$. The value of the energy of the HOS, $E_H$, is likewise close to the bulk Fermi energy $E_F$.

Note also that the Coulomb term µ* can be included in the usual way. It is worth noting that, unlike atoms, the positive charge in clusters is distributed over the cluster volume. As a result, for large clusters ($N \gtrsim 10^2$) the screening picture is similar to that in a bulk sample. In addition, because of discrete energy spectrum, we do not encounter a strong logarithmic singularity, but a threshold phenomenon, so that even at low $T_c$ the value of the coupling constant should exceed some critical value. Below we focus on the opposite case when the value of $T_c$ is large $(2\pi T_c / \tilde{\Omega} \gtrsim 1)$.

IV. **Critical temperature**

A. Theory

Based on Eq. (1), one can evaluate the critical temperature. At $T = T_C$ one should put $\Delta = 0$ in the denominator of the expression (1''), obtaining

$$\Delta(\omega_n) Z = \eta \frac{T}{2V} \sum_{\omega_{n'}} \sum_s \frac{\tilde{\Omega}^2}{\tilde{\Omega}^2 + (\omega_n - \omega_{n'})^2} \cdot \frac{\Delta(\omega_{n'})}{\omega_{n'}^2 + \xi_s^2} \qquad (5)$$

Note that the presence of the renormalization function $Z$ removes the divergence at $\omega_{n'} = \omega_n$.

The value of the parameter $\eta$ is close to its bulk value $\eta_b$. Indeed, the surface of the cluster can be treated as a scatterer (cf. [43]) and therefore the pairing is analogous to that in the case of a "dirty" superconductor analyzed in [5] , see also [44], whereby the mean free path is much shorter than the coherence length. Then the average value of $I^2$ is not affected by the scattering and one indeed finds that $\eta \approx \eta_b$ where $\eta_b$ is the bulk Hopfield parameter (see, e.g.,[40]). Note also that the characteristic vibrational frequency is close to the bulk value because pairing is mediated mainly by the short-wavelength part of the vibrational spectrum. Then Eq. (5) can be written in the form

$$\Delta(\omega_n)Z = \lambda_b \frac{T}{2\nu_b V} \sum_{\omega_{n'}} \sum_s \frac{\tilde{\Omega}^2}{\tilde{\Omega}^2 + (\omega_n - \omega_{n'})^2} \cdot \frac{\Delta(\omega_{n'})}{\omega_{n'}^2 + \xi_s^2} \Big|_{T_c} \qquad (6)$$

Here $\lambda_b = \eta \nu_b$ is the bulk coupling constant [45], $\nu_b = m^* p_F / 2\pi^2$ is the bulk density of states. With increasing cluster size the spectrum eventually becomes continuous. Then the integration over $\xi$ leads to the usual 3D equation for $T_c$ (see, e.g.,[37]).

Eq. (6) can be written in the dimensionless form

$$\phi_n = \sum_{n'} K_{nn'} \phi_{n'} \qquad (7)$$

where

$$K_{nn'} = g\tau_c \sum_s \left\{ \left[1 + (\tilde{\omega}_n - \tilde{\omega}_{n'})^2\right]^{-1} - \delta_{nn'} Z \right\} \left(\tilde{\omega}_{n'}^2 + \tilde{\xi}^2\right)^{-1} \Big|_{T_c} \qquad (8)$$

Here $\phi_n = \Delta(\omega_n)\tilde{\Omega}^{-1}$, $\tilde{\omega}_n = \omega_n \tilde{\Omega}^{-1}$, $\tilde{\xi}_j = \xi_j \tilde{\Omega}^{-1}$, $\tau = 2\pi T \tilde{\Omega}^{-1}$, and

$$g = \lambda_b \left(4\pi\tilde{\Omega}\nu_b V\right)^{-1} \qquad (8')$$

This expression for g is valid for neutral clusters as well as for ions. For neutral clusters one can also write $g = \lambda_b \left(6\pi \tilde{E}_F / N\right)$, $\tilde{E}_F = E_F \tilde{\Omega}^{-1}$, $E_F = p_F^2 / 2m^*$. We are not considering the Friedel oscillations of carrier density. For the relatively large clusters of interest their amplitude is comparatively small (see, e.g. [27]) and would modify the results only slightly.

The value of the critical temperature can be obtained from the matrix equation (cf. [35,36]):

$$\det|1 - K_{nn'}| = 0 \tag{9}$$

The expression for the kernel $K_{nn'}$ directly follows from Eq. (8), see Eq. (10) below.

Eq. (9) has a matrix structure. For the examples we considered (see below) the convergence was good even for 2x2 matrix, although we performed calculations with a higher accuracy (4x4) as well. Let us consider two different cases: (1) "magic" (spherical) clusters, and (2) open-shells (deformed) clusters.

B. "Magic" clusters

"Magic" clusters contain filled shells and are spherical in shape. As was mentioned above, in this case one can substitute $\sum_s \to \sum_j G_j$, $G_j$ is the degeneracy of the $j_{th}$ shell. The critical temperature can be evaluated from the matrix equation (9). With the use of Eqs. (2) and (8), we obtain

$$K^c_{nn'} = g\tau_c \sum_{j, n \neq n'} G_j \left( f^-_{n;n'} + f^+_{n;n'+1} \right) \chi_{n';\xi_j} \tag{10}$$

$$K^c_{nn} = g\tau_c \sum_j G_j \Big\{ 0.5 x_{n;0.5} x_{n;\xi_j} - (n+0.5)^{-1} \times$$

$$\tau_c^2 \sum_{m \neq n} \left[ (m+n+1)^2 - (n-m)^2 \right] f^-_{n;m} f^+_{n;m+1} (m+0.5) x_{m;\xi_j} \Big\}$$

Here

$$f^{\pm}_{n;r} = \left[ 1 + (n \pm r)^2 \tau_c^2 \right]^{-1} \tag{11}$$

$$x_{n;v} = \left[ (n+0.5)^2 \tau_c^2 + v^2 \right]^{-1} \tag{11'}$$

We focus on the case when $\tau_c \equiv (2\pi T_c / \tilde{\Omega}) \gtrsim 1$. In this case the matrix equation (9) converges rapidly (see below Eq. (14), and Sec. VI). Note also that the main contribution to the sums over shells in Eqs. (9) and (10) comes usually from the highest occupied shell (HOS) and the lowest unoccupied shell (LUS), so that with sufficient accuracy one can consider two terms (j≡H,L; $G_H = 2(2l_H+1)$; $G_L = 2(2l_L+1)$, $l_H$ and $l_L$ are the corresponding angular momenta). Let us introduce also the parameter $\tilde{\mu}$, defined by the relation:

$$\mu = E_H + \tilde{\mu}(E_L - E_H) \tag{12}$$

$$\xi_H = -\tilde{\mu}\Delta E, \quad \xi_L = (1-\tilde{\mu})\Delta E \tag{12'}$$

Here $\Delta E \equiv \Delta E_{LS} = E_L - E_H$. The chemical potential $\mu$ and, correspondingly, the parameter $\tilde{\mu}$ are determined by the relation (4) which for "magic" clusters has the form

$$N = \sum_j G_j \left( u_j^2 \varphi_j^- + v_j^2 \varphi_j^+ \right)$$

$$u_j^2, v_j^2 = 0.5 \left( 1 \mp \xi_j / \varepsilon_j \right) \tag{13}$$

$$\varphi_j^\mp = \left[ 1 + \exp(\mp \varepsilon_j / T) \right]$$

One can see from Eqs. (8) – (11), that the critical temperature $T_c$ is determined by parameters which can be measured experimentally. These parameters are: the number of valence electrons N, the energy spacing $\Delta E = E_L - E_H$. The magnitude of $T_c$ for a given nanocluster depends on these parameters and on values of $\lambda_b$, $E_F$ and $\tilde{\Omega}$, which are known for each material. The value $\Delta E$ has been calculated in different models (see, e.g., [16,17,27], but it also can be measured experimentally. As for the degeneracies $G_{H(L)} = 2(2l_{H(L)}+1)$, they can be obtained from symmetry considerations; they are similar in different models, e.g., "potential box" model, or jellium [27]. Consequently, our analysis employs parameters which can be determined from experimental data.

Below, we will describe a detailed calculation of $T_c$ for several specific clusters (see Sec. VI). Let us demonstrate first that for perfectly realistic values of the parameters a high value of $T_c$ can be obtained. Consider a cluster with the following parameter values:

$\Delta E = 65$ meV, $\tilde{\Omega} = 25$ meV, $m^* = m_e$, $k_F = 1.5 \times 10^8 \text{cm}^{-1}$, $\lambda_b = 0.4$,

the radius R = 7.5Å, and $G_H + G_L = 48$ (e.g., $l_H = 7$, $l_L = 4$); (14)

At this point, Eqs. (9) – (13) can be used. To estimate the value of $T_C$, one can use with good accuracy, the equation $1-K_{00}=0$ ( a more exact treatment, based on a 4x4 matrix changes the result only by several percent). One can also neglect by the relatively small dependence of $\mu$ on T, so that $|\xi_H| \cong |\xi_L| = \Delta E/2$. As a result, we are faced with the equation (see Eq.(10)):

$$1 = g_{eff.} F(\tau_c; \Delta \tilde{E})$$
$$g_{eff.} = 8g(G_H + G_L)$$
(15)

where g is defined by Eq. (8'), $\tau_c = 2\pi T_c/\Omega$, and

$$F(\tau_c; \Delta \tilde{E}) = \tau_c (\tau_c^2 + 1)^{-1} [(\tau_c^2 + (\Delta \tilde{E})^2)^{-1} - 4.5((4\tau_c^2 + 1)^{-1}(\tau_c^2 + (\Delta \tilde{E})^2)^{-1}]$$

Substituting the parameters values from above, we find by solving Eq.(15) that $T_C \cong 10^2 K(!)$.

One can see directly from Eq. (15) that the high degeneracies of the highest occupied shell (HOS) and the lowest unoccupied shell (LUS) play a very important role. Qualitatively, these degeneracies increase the effective electron-vibrational coupling $g_{eff}$, and more specifically, the effective density of states. As was noted in Sec.II, a sharp peak in the density of states at the Fermi level is very beneficial for pairing.

Consider another case with different values of the parameters:

$\Delta E = 0.1 eV$, $\tilde{\Omega} = 25$ meV, $m^* = 0.75 m_e$,

$k_F = 1.5 \times 10^8 cm^{-1}$, $\lambda_b = 0.5$, the radius R = 6A,

and $G_H + G_L = 48$ (e.g., $l_H=7$, $l_L=4$); then $g \cong 0.2$. (16)

Solution of Eqs.(9),(10) for the 4x4 matrix leads to $T_c \cong 120K$. In the first approximation one can use Eq.(15), and we obtain $T_c \cong 110K$.

The value of $T_C$ is sensitive to the magnitude of the HOS – LUS spacing. Indeed, if we calculate $T_c$ for a model cluster with parameters (16), and modify only $\Delta E = 0.1 eV \rightarrow 0.65$ meV, we obtain higher value of $T_c \cong 160K$ (!).

As was mentioned above, the value of $T_c$ also depends on the degeneracies $G_H=2(2l_H+1)$ and $G_L=2(2l_L+1)$, and they can be obtained from symmetry consideration. Indeed, in this section we focus on "magic" clusters which have a spherical shape (the case of deformed cluster will be discussed below, Sec. IVc) and for which the spectrum is determined by the "radial" quantum number $n$ and the orbital momentum $l$.

According to various experimental data and theoretical calculations (see e.g., the reviews [16,17]), the shell energy levels are not equidistant and the spacing between them varies considerably. As mentioned above, the most favorable case corresponds to a small $\Delta E$ and large $G_H$ and $G_L$. The value of the $\Delta E$ may be measured experimentally or calculated. While the sizes corresponding to "magic" numbers can be determined by mass spectrometry, photoemission or ionization potential measurements appear to be the best technique to determine the electron energy spectrum.

Theoretically, shell structure has been studied with the use of such models as the jellium model, a potential box, the Woods-Saxon potential, the oscillator model, etc. Note that the jellium model (see, e.g. [27]) as well as the spherical potential box model give similar sets of electronic shells and "magic" numbers. It is important that all the experimentally observed "magic" numbers such as $N_m$=2, 8, 32,132, 198, etc. are among those obtained by calculations based on these models; this offers strong support for the classification based on their use. One should note, however, that the models also predict some "magic" numbers which have not been observed. This is due to the fact, that some of the HOS and LUS are anomalously close, and their hybridization leads to the formation of an incomplete single shell and to Jahn-Teller distortion. In Sec.VI below we describe a detailed calculation for some specific clusters, but first let us consider the case of a cluster with an unfilled electronic shell.

## C. Incomplete shells

Clusters with partially occupied shells undergo a Jahn-Teller transition manifested as a shape distortion. The transformation from spherical to ellipsoidal shape splits the degenerate levels. The scale of the splitting depends on the number of electrons removed from the HOS (or added to LUS) and on the properties of the material (see below, Eq. (16) and Sec. VI). Because of the splitting, one should not use Eqs. (10), (11), (13) which are valid only for the "magic" clusters. The value of $T_C$ for clusters with incomplete shells can be calculated (see below, Sec. VI) with the use of Eq. (9) and more general expressions obtained from Eqs. (10-13) with the replacement $\sum_j G_j \to \sum_s$, i.e., a summation over all levels formed by the splitting of highest occupied and lowest unoccupied shells. The density of states is now spread over different energy levels, and such weakening of the sharp peak feature is not a positive factor for $T_C$. On the other hand, the removal of electrons from HOS strongly affects the position of the chemical potential, and this factor turns out to be favorable for pairing. The best scenario would correspond to clusters with almost filled shells (e.g., $N=N_m-2$, $N_m$ is a "magic" number) and, correspondingly, a small deviation from sphericity. In this case the HOS turns into a set of close levels classified by the projection of their angular momentum *m*. The picture of splitting is similar to that in atomic nuclei (cf., e.g., [46]). To calculate the magnitude of the splitting, one can use the following expression [47]:

$$\delta E_l^m = 2 E_l^{(o)} U r_l^m \qquad (17)$$

$$r_l^m = 2[l(l+1) - 3m^2](2l+3)^{-1}(2l-1)^{-1}$$

Here U is the deformation parameter, $E_l^{(o)} = E_H \equiv E_{HOS}$. An explicit expression for the deformation parameter can be found [48] by minimizing the total energy $\delta E = \delta E_{el} + \delta E_{def.}$, where $\delta E_{el}$ is described by Eq. (17) and the deformation contribution $\delta E_{def.} = 3U^2 V c_{el}$, $c_{el} = c_{11} - c_{12}$; $c_{11}$, $c_{12}$, are the elastic constant, and V is the cluster volume. The deformation parameter is determined by the condition $\partial(\delta E)/\partial U = 0$. The scale of splitting is different for various materials. We will discuss this question in detail in Sec. VI. Let us estimate the scale of the splitting for the model cluster, see above, Sec. IV.B. Assume $c_{el} = 10^{12}$ dynes·cm$^{-2}$, R = 7.5Å, $l=7$, R is the cluster radius. The value of the energy $E_H$ is on order of $E_F = 8.5$eV. Then Eq. (17) yields $\delta E \cong 10$meV. Such a splitting is relatively small. It is worth noting that there is a realistic situation with even much smaller splitting. It can arise in the case of an overlap of the manifolds of sub-levels formed by splitting of HOS and LUS. For example, for the cluster with initial spacing $\Delta E^M \approx 0.1$ eV (here the superscript "M" stands for the initial "magic" cluster with N=168), $g_{eff,}=10$ (see above, Sec.IVc), R=6A, $c_{el}=10^{12}$dynes/cm$^2$, and N=166 (two electrons removed from HOS), there is an overlap of the manifolds with $l=7$ (HOS) and $l=4$ (LUS). A direct calculation gives

ΔE≈2 meV for the smallest excitation energy. The removal of two electrons does not noticeably affect the critical temperature which is close to that for the cluster with N=168 ($T_c$≈110K). However, the pairing drastically affect the energy spectrum (see below, Sec. VA).

## V. Energy spectrum

### A. Ground State

The onset of pairing has a great impact on the cluster energy spectrum. Let us begin by evaluating the gap parameter and, correspondingly, the spectrum at T=0K. We start with Eq. (1)), whose solution at T=0K yields the spectrum. Indeed, it is seen directly from Eqs. (3) and (4) that the spectrum has the form (4''), where the gap parameter is the root of the equation

$$\varepsilon_{0;s} = \Delta\left[i(\xi_s^2 + \varepsilon_{0;s}^2)^{1/2}\right] \tag{18}$$

Eq. (1) at T=0K can be written in the following dimensionless form:

$$\phi(x) = g\sum_s \int dx_1 L(x;x_1) \frac{\phi(x_1)}{x_1^2 + \tilde{\xi}_s^2 + \phi^2(x_1)} \tag{19}$$

Here $L(x;x_1) = f^- + f^+ - 4x_1^2 f^- f^+$; $f^\pm = \left[1+(x\pm x_1)^2\right]^{-1}$; $\phi(x) = \Delta(x)\tilde{\Omega}^{-1}$; $x = \omega\tilde{\Omega}^{-1}$. The solution of Eq. (19) can be sought in the form:

$$\phi(x) = \beta(1+\alpha x^2)^{-1} \tag{20}$$

The position of the chemical potential μ [or the parameter $\tilde{\mu}$, see Eqs. (12), (12')] is determined by Eqs. (3), (4). Note that the value of $\tilde{\mu}$ is strongly affected by the pairing and must be determined in a self-consistent way. For example, for a "magic" cluster in the absence of pairing $\tilde{\mu} = 0.5$ at T=0K (see Appendix B). This changes if $\phi \neq 0$ (see below). The values of the parameters α, β, and $\tilde{\mu}$ can be obtained from Eqs. (19), (3), and (4) by an iterative procedure, namely by minimizing the quantity $<\phi^{(1)} - \phi^{(2)}>/<\phi^{(2)}>$, where $\phi^{(1)}$ and $\phi^{(2)}$ are the first and second iterations. Subsequently, the gap parameter can be evaluated from Eqs. (18), (20) by solving the corresponding cubic equation. The solution has the following form:

$$\varepsilon_{0;s} = (2/\sqrt{3})\eta_s \sin\left\{(1/3)\arcsin\left[(3\sqrt{3}/2)(\beta/\alpha\eta_s^3)\right]\right\} \qquad (21)$$

Here

$$\eta_s = \left[\alpha^{-1} - \xi_s^2\right]^{1/2} \qquad (21')$$

For "magic" (spherical) clusters the major contribution comes from the HOS and LUS shells. The method described above, allows one to calculate the gap parameters $\varepsilon_{0;H}$ and $\varepsilon_{0;L}$. The minimum excitation energy is equal

$$\Delta\varepsilon = \varepsilon_H + \varepsilon_L \qquad (22)$$

where

$$\varepsilon_H = \left[\tilde{\mu}^2(\Delta E)^2 + \varepsilon_{0;H}^2\right]^{1/2} \qquad (22')$$

$$\varepsilon_L = \left[(1-\tilde{\mu})^2(\Delta E)^2 + \varepsilon_{0;L}^2\right]^{1/2} \qquad (22'')$$

where $\varepsilon_{0;H}$ and $\varepsilon_{0;L}$ are determined by Eqs. (21) and (21') with $\eta_{H;L} = [\alpha^{-1} - \xi_{H;L}^2]^{1/2}$, $\xi_H$ and $\xi_L$ are the first terms on the right-hand side of Eq. (4'').

For example, for a cluster with the following realistic set of parameters : $\Delta E$=65 meV, $E_F$=8 eV, $\tilde{\Omega}$=25 meV, $\lambda_b = 0.4$ we obtain: $\varepsilon_H$=50 meV, $\varepsilon_L$=-30 meV, so that $\Delta\varepsilon$=80 meV. This value noticeably exceeds the shell spacing in the absence of the pairing, $\Delta E$= 65 meV.

The spectral changes induced by the pairing are much larger for clusters with slightly incomplete shells. As was discussed in Sec. II and IV.C , in this case the HOS and LUS are split (Fig. 1 ). This case can be treated with the help of Eqs. (3), (17), (19); however, in this case one should take into account a number of separate terms "s", corresponding to different values of *m*. The general picture is that the upper occupied level is always incompletely filled, and the minimum excitation energy in the absence of pairing is rather small. Especially interesting is the case when split HOS and LUS manifolds overlap. For example, for the case discussed at the end of Sec.IVc the interval $\Delta E$ was ≈2 meV. With pairing, we obtain $\Delta\varepsilon$≈40 meV, a dramatically altered value. Therefore, $\Delta\varepsilon$>>$\Delta E$, and pairing is seen to lead to a large increase in the minimum excitation energy.

## B. Region near $T_C$. Ginzburg-Landau functional

Let us consider the pairing at temperatures close to $T_C$. Once again, we can employ the general equations (1)- (3). The order parameter $\Delta \to 0$ as $T \to T_C$. As a result $\Delta(\omega_n) \ll \omega_n$, and the right–hand side of Eq. (1) can be expanded in a series in terms of $\Delta^2$. We can start with the expression (at $T \sim T_C$) for the solution of Eq.(1) similar to Eq. (20).

$$\phi_n = b(1 + a\tilde{\omega}_n^2)^{-1} \tag{23}$$

Here $\phi_n = \Delta(\omega_n)\tilde{\Omega}_n^{-1}$, $\tilde{\omega}_n = \omega_n \tilde{\Omega}^{-1}$, and $b^2 = \sigma\delta\tau \equiv \sigma\tau_c(1-t)$, t=T/$T_C$. Near $T_c$ the expression (23) also can be written as (for small $n$): $\phi_n = b(1 + a\tau_c^2/4)^{-1} f_n$; here $f_n = \{1, f_1, \ldots\}$. The amplitudes $f_n$ as well as the parameter "$\sigma$" could be calculated from Eqs. (1), (4). More specifically, one can use Eq. (7), but the kernel $K_{nn'}$ should be written as a power series in of (1-t). As a result, near $T_c$ the kernel can be written as a sum

$$K_{nn'} = K^c_{nn'} + K^{(1)}_{nn'} \tag{24}$$

where $K^c_{nn'}$ is described by Eq.(10), and

$$K^{(1)}_{nn'} = g\delta\tau \sum_{\substack{S \\ n \neq n'}} \left[ (1 - \tau_c(n-n')^2) f^+_{n;n'} + (1 - \tau_c^2(n+n'+1)^2) f^+_{n;n'+1} \right] \chi_{n';\tilde{\xi}_s} -$$

$$-2(f^-_{n;n'} + f^+_{n;n'+1})(\tau_c^2(n'+0.5)^2 + \tilde{\xi}_s S_1/S) \chi^2_{n';\tilde{\xi}_s} -$$

$$-g\tau_c \sum_S (f^-_{n;n'} + f^+_{n;n'+1})(\phi^2_{n'} + \tilde{\xi}_s(S_2/S)\beta^2) \chi^2_{n';\tilde{\xi}_s}$$

$$K_{nn}^{(1)} = 2g\delta\tau \sum_S \left(1 - 4\tau_c^2(n+0.5)^2\right) f_{2n;1}^+ \chi_{n';\tilde{\xi}_s} -$$

$$-2\left(\tau_c^2(n+0.5)^2 + \tilde{\xi}_s S_1/S\right) f_{2n;1}^+ \chi_{n;\tilde{\xi}_s}^2 -$$

$$-2g\tau_c \sum_S \left(\phi_n^2 + \tilde{\xi}_s (S_2/S)\beta^2\right) f_{2n;1}^+ \chi_{n;\tilde{\xi}_s}$$

where $f_{n;r}^{\pm}, \chi_{n;v}$ are defined by Eqs.(11),(11'), and

$$S = \sum_s (2+\varphi_s^+)^{-1}; \quad S_1 = \sum_s \tilde{\xi}_s (2+\varphi_s^+)^{-1} \tag{25}$$

$$S_2 = \sum_s (1-\alpha\tilde{\xi}_s^2)^{-2} \left\{\tau_c \varphi_s^- / 4\pi \left|\tilde{\xi}_s\right| - \tilde{\xi}_s^{-1}\right\} (2+\varphi_s^-)^{-1}$$

$$\varphi_s^{\pm} = \exp\left(\left|\tilde{\xi}_s\right|/T_C\right) \pm \exp\left(-\left|\tilde{\xi}_s\right|/T_C\right) \quad \xi_s = E_s - \mu_c; \quad \mu_c \equiv \mu(T_c)$$

One can write the following expression for the thermodynamic potential:

$$\delta\tilde{\Theta}_s = \tilde{A}\left[-(\tau_c - \tau)b^2 + (2\sigma)^{-1}b^4\right] \tag{26}$$

where $\delta\Theta_s = \Theta_s - \Theta_n$ is the change in the thermodynamic potential caused by the transition to the superconducting pairing state, $\tilde{\Theta}_s = \Theta_s \tilde{\Omega}^{-1}$. Eq. (26) is analogous to the Ginzburg-Landau functional (cf., e.g., [49]). Indeed, the condition $\partial\Theta/\partial b^2 = 0$ leads to the relation $b^2 = \sigma\delta\tau$.

The parameter $\tilde{A} = -b^{-2}(\partial\Theta_s/\partial\tau_c)$, see Eq. (26), can be evaluated with use of the well-known expression (see, e.g., [37]):

$$\left(\partial\tilde{\Theta}_s/\partial\lambda_{eH}\right) = \lambda_{eff}^{-1} <H_{inf}> \tag{27}$$

$\lambda_{eff} = \eta/v$, see Eq. (1). With the relation $\left(\partial\tilde{\Theta}_s/\partial\lambda_{eff}\right) = \left(\partial\tilde{\Theta}_s/\partial\tau_c\right)\left(\partial\tau_c/\partial\lambda_{eff}\right)$ we obtain:

$$\tilde{A} = \tau_c^2 \tilde{s} / \pi (\partial \tau_c / \partial g) \qquad (28)$$

Here g is defined by Eq. (8´), and

$$\tilde{s} = 0.25 \sum_{\substack{s,s' \\ n,n' \geq 0}} \left( f^+_{n;n+1} + f^-_{n;n'} \right) x_{n;\xi_s} x_{n';\xi_{s'}} \phi_n^{ext.} \phi_{n'}^{ext.} \qquad (29)$$

The quantities $f^\pm_{n,r}$ and $x_{n;\nu}$ are defined by Eqs. (11), (11'), and $\phi_n^{ext.}$ is the general expression for the order parameter, see Appendix, Eq. (A.2).

Eqs. (1), (25), (28) allows us to calculate the major parameters such as σ, a, A, which describe the pairing near $T_c$. As before, these values are not universal and depend on the properties of the materials. For example, for "magic" clusters the values depend on the degeneracies $G_H$ and $G_L$, the spacing $\Delta E$, etc. For the model case considered above, Eq.(14) the calculations leads to: δ=0.5 , $\tilde{A} \approx 16$ , a=0.1 .Below, these parameters will be calculated for various specific systems, see Sec. VI.

### C. Fluctuations

Eq. (26) allows one to investigate the interesting subject of fluctuations. In general, it is known (see, e.g. [50]) that a decrease in size increases the role of fluctuations. In connection with this, it is important to note that the large values of $T_C$ and of the gap parameter for the clusters studied here result in a relatively small coherence length ,comparable with the cluster size. The situation is similar to that in the high $T_C$ oxides.

Let us estimate the broadening of the transition $\delta T_c/T_c$ due to fluctuations. This calculation can be performed with the use of Eq. (26), cf. [49 ,Ch.8]. First, one calculates $\delta\tilde{\Theta}_{s;min}$ which is equal to $\delta\tilde{\Theta}_{s;min} = -(\tilde{A}^2 \sigma \tau_s / 2)(1-t)$. The width of the transition is determined by the condition $\delta\Omega \sim kT$ (see, e.g., [49]). As a result, we arrive at the following expression:

$$\frac{\delta T_C}{T_C} \approx (2\pi)^{-1} \left(\tilde{\Omega}/2\tilde{A}\sigma T_C\right)^{1/2} \qquad (30)$$

where $\tilde{A}$ is defined by Eq. (28). A direct calculation for the case specified in Eq. (16) shows that the transition broadening is on the order of $(\delta T_C/T_C) \approx 5\%$. Similar values are obtained from various specific clusters (see below, Sec. VI). A width of the magnitude noticeably exceeds that for bulk superconductors, but is still relatively small.

VI. **Specific clusters**

In the previous section we demonstrated that the cluster with realistic parameters , Eq. (14), possesses a $T_c \approx 10^2 K$, which greatly exceeds that for conventional metals. The presence of shell structure and the resulting large degeneracy are key factor leading to such high $T_c$. Based on the general method described above, one can calculate the critical temperature and the gap parameter for some specific clusters. We will consider "magic" and deformed clusters separately.

A. "Magic" clusters

The calculation of $T_C$ is based on Eqs. (8) –(13). As was noted above (see the discussion after Eq.(13), its value depends on the specific parameters for a given cluster. The general method described above (Secs.III-V) can be employed to analyze clusters of different materials. Let us consider some examples. In this section we describe calculations performed for certain Ga, Al, Zn, and Cd clusters. This choice is not accidental. The important facts are that shell structure has been observed experimentally in all these clusters [20-25], and at the same time these materials are superconducting in the bulk state.

We performed calculations (see Eq. (9)) with 4x4 matrix, giving a high numerical accuracy (< 1%). Note that even 2x2 matrices provide adequate accuracy (<10%). It turns out that the major contribution comes from H and L shells, so that in the summation over j in Eqs.(9), (10), and(13) it is sufficient to retain just the terms with $G_H$ and $G_L$.

We selected two "magic" numbers: $N_m$=168 and $N_m$=380. Indeed, such clusters are characterized by large values of orbital momentum $l$ and by small $\Delta E$. Let us consider first the $Al_{56}$ cluster (each Al atom contains 3 valence electrons, so that N =168). The highest occupied shell corresponds to $l=7, n=1$ and contains 30 electrons. In addition to various materials parameters (see below) one needs to know the value of $\Delta E = E_L - E_H$. The energy $E_L$ corresponds to the lowest excited level for the selected "magic" cluster $N_m$ (here $N_m=168$). Strictly speaking, this $E_L$ is different from the $E_H$ for the next "magic" cluster $N_{m'}=186$, since these clusters differ in size. Nevertheless, this difference is small (in the "potential box" model it is of order of several percent, see also [51]).

The spacing $\Delta E$ can be measured experimentally, e.g., as the difference in the ionization potentials for the two neighboring "magic" clusters. According to [20] the Al cluster with N=168 is, indeed, "magic" (a sharp drop of the ionization potential is observed). Theoretically, the next cluster with complete shell corresponds to $N_{m'}=186$. However, according to [20], the next "magic" number corresponds not to N=186, but rather to N = 198. This is probably due to the closeness and consequent hybridization of the shells with $l=4,n=2$; $l=2,n=3$; $l=0,n=4$. As a result, the next spherical shape will correspond to N = 198 with total number of LUS states is $G_L=30$ so that $G_H+G_L=60$. According to [20], $\Delta E \approx 0.1 eV$. With the use of these data and the parameters for $Al_{56}$ clusters (R≈6.5A, $\tilde{\Omega} \approx 350K$, $\lambda_b \approx 0.4$ [52], $m^* \approx 1.4 m_e$, $k_F = 1.75 \cdot 10^8 cm^{-1}$) we obtain $T_c \approx 90K$.

As the next example consider the $Ga_{56}$ cluster, which is similar to $Al_{56}$ (the Ga atom also has 3 valence electrons). Because of the similarity in the electronic structure and close values of the work functions W ($W_{Al} \approx 4.3 eV$, $W_{Ga} \approx 4.2 eV$), one can assume that the values of the ionization potentials are also close. We can use the values: $\tilde{\Omega} \approx 270 K$, $\lambda_b \approx 0.4$ [52], $m^* \approx 0.6 m_e$, $k_F = 1.7 \cdot 10^8 cm^{-1}$. Estimating the value of $T_c$ with use of Eq. (15), we obtain: $T_c \approx 170 K$. A more accurate calculation based on Eq. (9)-(13) leads to $T_c \approx 145 K(!)$ which greatly exceeds the bulk value ($T_c^b \approx 1.1 K$). As emphasized above, such a drastic increase is due to the large degeneracy and, correspondingly, to the large effective density of states at the Fermi level. Another important factor is the relatively small value of the interval $\Delta E$.

Let us turn to a calculation of $T_c$ for other clusters. In the absence of experimental measurements of $\Delta E$, the calculation can be carried out in the following way. In order to illustrate the method, lets consider ,at first, the same $Ga_{56}$ cluster. Let us write $\Delta E = E_H \gamma$, i.e., ($\gamma = (E_L/E_H) - 1$). For relatively large clusters ($N \gtrsim 10^2$) $E_H \approx E_F$. The parameter $\gamma$ can be estimated from the "potential box" model. As a result, we have $\Delta E \cong E_H \gamma$, $\gamma \cong (E_{H;m'}/E_{H;m}) - 1$. The parameter $\gamma$ can be written as $\gamma = (Z_{m'}/Z_m)^2 - 1$, $z_m$, $Z_{m'}$ are zeros of the Bessel function $J_{e+1/2}(x)$. In our case with $N_m = 168$ and $N_{m'} = 186$, $\gamma \approx 4 \cdot 10^{-3}$. After a calculation with Esq. (8) – (13) , we obtain for the $Ga_{56}$ clusters $T_c \approx$ 132K, which is close but lower than the value obtained above. One can see that such a model provides a low limit of $T_C$.

Consider now $Zn_{84}$ and $Cd_{84}$ clusters (where each atom contains two valence electrons). Using the following parameters [52] : $\tilde{\Omega}=275K, \lambda_b \approx 0.4, E_F \approx 12eV$ (for Zn) and : $\tilde{\Omega}=210K, \lambda_b \approx 0.4, E_F \approx 10eV$ (for Cd), we obtain, after calculations $T_C \approx 95K$ (for Zn) and $T_C \approx 65K$ (for Cd). These values also greatly exceed those for the bulk metals ($T_C^b \approx 0.9K$ for Zn and $T_C^b \approx 0.6K$ for Cd.).

Another promising "magic" number is $N_m=380$. HOS corresponds to $l=10$, $n=1$, so that the HOS contains 42 electrons, and the LUS has $l=4$, $n=3$. To best of our knowledge, the measurements of the HOS-LUS interval have not been performed for such systems. For an estimate we use the "potential box" model; the values of $\Delta E$ for $Zn_{190}$ and $Cd_{190}$ clusters appear to be close: $\Delta E = 6$ meV. A calculation performed with use of Eqs. (8) –(13) and a 4x4 matrix leads to the following: $T_c \approx 105K$ for the $Zn_{190}$ clusters and $T_c \approx 90$ K for the $Cd_{190}$ clusters.

B. Incomplete shells.

Clusters with incomplete shells undergo a Jahn-Teller distortion and acquire non-spherical, ellipsoidal shapes. We consider clusters with slightly incomplete shells. As noted above, in this case one may expect small shape deformation, and the energy spectrum can be viewed as a small splitting of the initially degenerate HOS and LUS. For example, clusters with N=166 can be treated as nanoparticles whose electronic system is obtained by the removal of two electrons from the "magic" structure with N=168. The HOS of such a deformed cluster is split, with 28 electrons occupying levels corresponding to different values of the projection of the orbital momentum, so that $|m| \leq l=7$. Here we focus on clusters with an even number of electrons (odd-even effects will be discussed below, Sec.VII).

The scale of the splitting depends on the number of vacancies in the shell (the number of "removed" electrons), and on the elastic parameter $c_{el}$ (see Sec.IV c). Fortunately, for Zn and Cd clusters, the latter parameter is large (see below) and the splitting should be small.

Consider the $Zn_{83}$ cluster (N=166). The HOS of the deformed cluster is made up of 8 sublevels ($|m|\leq l$), each with a degeneracy $G_s=4$, except $m=0$ with $G_0=2$. The spectrum can be calculated from Eq. (17); the parameter $c_{el}=1.4 \times 10^{12}$ dynes/sm$^2$ [53]. The upper level ($|m|=7$) of the incomplete shell contains only two electrons. This factor strongly affects the position of the chemical potential, which is especially important at T=0K (see below). The critical temperature can be calculated from the general equations (9)-(13) with the substitution $\sum_j G_j \to \sum_s$, that is, by summation over all split levels. As a result, we obtain $T_c \approx 106K$ for the $Zn_{83}$ cluster. A similar analysis for the $Cd_{83}$ cluster leads to the value $T_c \approx 85K$. These values of $T_C$ are higher than for the "magic" clusters with N=168; this is due to the change in the position of the chemical potential.

As mentioned, pairing results in a considerable impact on the cluster electronic spectrum. It turns out that this impact is especially strong for clusters with incomplete shells, and is manifested in the appearance of the pairing gap parameter. We will discuss this aspect in the next section.

### C. Energy gap.

The effect of pairing on the spectrum is much stronger for the clusters with slightly incomplete shells. Consider, e.g., the $Cd_{83}$ cluster (N=166). Because the uppermost level of the set formed by the splitting of its HOS is not fully occupied (a complete shell corresponds to N=168), the smallest excitation energy (in the absence of pairing) $\Delta E_{min}= E_{|m|=l} - E_{|m|=l-1}$ is not large. Indeed, based on Eq. (17), one can find $\Delta E_{min} \approx 6$ meV ($c_{el}=1.25 \times 10^{12}$ dyn/cm$^2$). Based on Eqs. (4), (18)-(20) one can calculate the gap parameters $\varepsilon_{0;L}$ and $\varepsilon_{0;H}$ and $\tilde{\mu}$. we obtain $\alpha \approx 2 \times 10^{-2}$, $\beta \approx 0.9$, and this leads to value $\Delta \varepsilon_{min;} \approx 34$ meV. Similar result can be obtained for $Zn_{83}$ clusters ($c_{el}=1.4 \times 10^{12}$ dyn/cm$^2$); $\alpha \approx 4.5 \times 10^{-2}$, $\beta \approx 0.9$. For these clusters $\Delta E_{min} \approx 6.5$ meV;

$\Delta\varepsilon_{min} \approx 42.5$meV. Therefore $\Delta\varepsilon_{min} >> \Delta E_{min}$ and, indeed, the impact of pairing is very significant.

VII. **Proposed experiments. Discussion.**

We have described the phenomenon of pair correlation in an *isolated* nanocluster. The question arises, how can this correlation be observed and what kind of experiment can verify its presence? Of course, if a tunneling network of such nanoclusters were built, a macroscopic superconducting current could be observed. We will discuss this aspect below, but first, let us address the possibility of observing pair correlation in an isolated cluster.

Pairing leads to a strong temperature dependence of the excitation spectrum. At $T>T_C$ the minimum excitation energy is given by $\Delta E_{min}$ ($\Delta E_{min} \equiv \Delta E^M = E_L - E_H$ for "magic" clusters, and $\Delta E_{min} = E^H_{|m|=l} - E^H_{|m|=l-1}$ for clusters with slightly incomplete shells); especially interesting case corresponds to overlap of HOS and LUS manifolds. Below $T_C$ and especially at low temperatures close to $T=0$K, the excitation energy is strongly modified by the gap parameter and noticeably exceeds that in the region T>$T_c$. The shift is especially dramatic for clusters with slightly unoccupied shells. We demonstrated above (Secs. IVC, VIB ) that for such clusters the ratio $\Delta\varepsilon_{min}/\Delta E_{min}$ can be ~ 6-7. An overlap of HOS and LUS manifolds leads to even greater values. A change of such magnitude in the excitation energy should be experimentally observable and would represent a strong manifestation of pair correlation. Generating beams of isolated metallic clusters at different temperatures (see, e.g., [4, 54]) in combination with mass selection , would allow one to focus on clusters of specific size at various temperatures. A measurement of the energy spectrum, in particular a determination of $\Delta E_{min}$ (for example, by photoelectron spectroscopy, see, e.g. [55, 56]) would reveal a strong temperature dependence of the spectrum. For example, in Ga clusters (N=168, $T_C \approx$ 130K) one should observe a large difference in $\Delta E_{min}$ at low temperature region and above $T_c \approx$ 130K. Similarly, for Cd clusters with N=166 a large difference should be observed between spectra at low temperatures and at T>$T_c \approx$85K. The use of Ga and Cd nanoclusters for such experiments looks reasonable, because these materials are superconducting and, as mentioned above, the existence of electronic shell structure in their clusters has been confirmed experimentally. An experiment of this type would be both realistic and informative. Note that pairing would manifest itself differently from a structural transition [57], since the former strongly affects the spectrum near the H and L shells, whereas the latter would modify the entire spectrum. If it is possible to place small nanoclusters into a tunneling barrier, then the spectrum can be determined with the use of inelastic tunneling spectroscopy similar to that employed in [1,2]. In this case there will be no problem related to optical selection rules.

As noted, clusters with pair correlation are promising building blocks for tunneling networks. Charge transfer between clusters, provided by Josephson coupling, would give rise to a macroscopic superconducting current at high temperatures. Such a network could be prepared by depositing clusters on a surface. It requires special methods of growing isolated clusters in a matrix without strong disturbance of their shapes and spectra. Another possibility has been considered in [7] (see above, Sec. II) and envisions a molecular (cluster) crystal where clusters form an ordered 3D lattice. There has been noticeable progress in this field, see, e.g., [58-61].

The pair correlation also can manifest itself in odd-even effects in cluster spectra. Such an effect has been observed in [2], but for much larger particles ($N \approx 10^4$-$10^5$). It would be interesting if it were possible to perform similar spectroscopy for small nanoclusters displaying shell structure such as, e.g., $Ga_{56}$ or $Cd_{83}$ studied here.

In this paper we studied simple metallic clusters (Al, Ga, Zn, Cd). In principle, one may consider a variety of systems, including those containing more complicated alloy clusters (see, e.g., [62]). In principle, it may be possible to raise $T_c$ even higher, possibly up to room temperatures. Indeed, one can see, e.g., from Eq.(15), that an increase in T$_c$ can be achieved by changing the parameters ($\tilde{\Omega}, \Delta E$, etc.) in the desired direction. The study of pair correlation and its impact on shell structure in small nanoclusters is an interesting and promising field.

**Aknowledgements**

The authors are very grateful to J. Friedel for many fruitful discussions. We are especially indebted to V.V. Kresin for useful discussions regarding the properties of nanoclusters and their spectroscopy. We would like to thank A.Barone, M.Beasley, R.Dynes, A.Goldman, T.Herrmannsdorfer, S.Wolf, and M.Tinkham. The research of VZK was supported by DARPA under Contract No. 05U716. The research of YNO was supported by the CRDF under Contract No. RP1-2565-MO-03 and by RFBR (Russia).

## Appendix A. Order parameter near $T_c$.

Near $T_c$ the general equation (1) for the order parameter can be linearized. This allows one to seek the solution at small $n$ in the more general form (cf. Eq.(23)):

$$\phi_n = b\left(1 + b_1 \tilde{\omega}_n^2\right)\left(1 + a\tilde{\omega}_n^2 + a_1 \tilde{\omega}_n^4\right)^{-1} \tag{A.1}$$

One should select the parameters, so that (A.1) form the best fit to the form

$$\phi_n = b\left(1 + b_1 \tau_c^2/4\right)^{-1}\left[1 + a(\tau_c^2/4) + a_1(\tau_c^2/4)^2\right]^{-1} \tag{A.2}$$

There should not be poles for the analytical continuation of (A.1) in the upper half-plane. Near $T_c$ the function $\phi_n$ can be represented in the form $\phi_n = \phi(f_n + \phi_n^2 \psi_n + \ldots)$ where $\psi_n$ corresponds to lower eigenvalues of $\hat{K}$. Using the condition $\langle f^+ \psi \rangle = 0$, one can obtain the following equation for the parameter $b$:

$$f_n^+ \hat{K}_{nn'} f_{n'} = 0 \tag{A.3}$$

where $\hat{K}_{nn'}$ is defined by Eqs. (24).

Eq. (4) allows us to determine the shift $\delta\mu = \mu(T) - \mu(T_c)$ and we obtain:

$$\delta\tilde{\mu} = -(S_1 + cS_2)^{t/S} \tag{A.4}$$

Here $\delta\tilde{\mu} = \delta\mu \tilde{\Omega}^{-1}$, $\delta\tau = \tau - \tau_c$, b, $S_1$, $S_2$ and S are defined by Eqs. (25), (A3).

## Appendix B. Dependence μ(T).

Let us evaluate the dependence $\tilde{\mu}(T)$, for the "magic" clusters in the absence of pairing. This dependence is influenced mainly by the contributions of two shells, HOS and LUS. At finite temperature we have

$$G_H\left[1+\exp(\tilde{\mu}\Delta E/T)\right]^{-1} = G_L\left[1+\exp((1-\tilde{\mu})\Delta E/T)\right]^{-1} \quad (A.5)$$

where $G_{H;L} = 2(2l_{H;L}+1)$, $\Delta E = E_L - E_H$. As a result we obtain

$$\tilde{\mu} = (T/\Delta E)\ln\left\{p + \left[p^2 + (G_H/G_L)\exp(\Delta E/T)\right]^{1/2}\right\} \quad (A.6)$$

$p = (l_H - l_L)/(2 l_L + 1)$.

Eq. (A.6) determine the dependence $\tilde{\mu}(T)$. At T=0K, we obtain $\tilde{\mu} = 0.5$, so that the chemical potential, indeed, is located in the middle of the interval $\Delta E$.

# References

1. D.Ralph, C.Black, M.Tinkham, Phys.Rev.Lett. 74, 3241 (1995); *ibid.* 76, 688(1996); *ibid.* 78, 408 (1997)
2. M.Tinkham, J.Hergenrother, J.Lu, Phys.Rev.B 51,12649(1995)
3. M. Strongin, R. Thompson, O. Kammerer, J. Crow. Phys. Rev. B1, 1078 (1970)
4. R. Moro, X. Xu, S. Yin, W. de Heer, Science 300, 1265 (2003)
5. P. Anderson, J. Phys. Chem. Solids, 11, 59 (1959)
6. W.Knight, in *Novel Superconductivity*, ed. by S.Wolf and V.Z.Kresin, (Plenum, New York, 1987), p. 47.
7. J. Friedel, J.Phys. 2, 959 (1992).
8. F.Braun and J. von Delft , Phys. Rev. B 59, 9527 (1999)
9. K.Matveev and A.Larkin, Phys. Rev. Lett. 78, 3749 (1997)
10. H.Heiselberg, Phys. Rev. A 63, 043606 (2001)
11. M.Baranco, E.Hernandez, R.Lombard, L.Serra, Z.Phys.D 22, 659 (1992)
12. H.Boyaci, Z.Gedik, I.Kulik,J. of Supercond. 14 ,133 (2001)
13. J. Perenboom, P. Wyder, and F. Meier, Phys. Reports 78, 173 (1981)
14. J. von Delft, D. Ralph, Phys. Rep. 345, 61 (2001)
15. Y. Ovchinnikov and V. Kresin, Eur. Phys. J. B 45, 5 (2005); 47, 333 (2005)
16. W. de Heer, Rev. Mod. Phys. 65, 611 (1993)
17. M.Brack, Reviews of Modern Physics, 65, 677 (1993)
18. V.V.Kresin and W.Knight, in *Pair Correlations in Many-Fermion Systems,* ed. by V.Z.Kresin (Plenum, New York, 1998), p.245
19. W. Knight, K. Clemenger, W. de Heer, W. Saunders, M. Chou, M. Cohen, Phys. Rev. Lett. 52, 2141 (1984)

20. K. Schriver, J. Persson, E. Honea, R. Whetten, Phys. Rev. Lett. 64, 2539 (1990); M. Pellarin, B. Baguenard, M. Broyer, J. Lerme, J. Vialle, A. Perez, J. Chem. Phys. 98, 944 (1993)

21. M.Pellarin, B.Baguenard, C.Bordas, M.Broyer, J.Lerme, J.Vialle, Phys.Rev.B 48, 17645 (1993); B.Baguenard, M.Pellarin, C.Bordas, J.Lerme, J.Vialle, M.Broyer, Chem. Phys. Lett. 205, 13 (1993)

23. I.Katakuse, T.Ichihara, Y.Fujita, T.Matsuo, T.Sakurai, Matsuda, Int. J. of Mass Spectrometry and Ion Processes 69, 109 (1986)

24. M.Ruppel, K.Rademann, Chem.Phys.Lett. 197, 280 (1992)

25. J. Lerme, P.Dugourd, R.Hudgins, M.Jarrold, Chem. Phys. Lett. 304, 19 (1999)

26. J.Martin, R.Car, J.Buttet, Surf.Sci. 106, 265 (1981)

27. W. Ekardt, Phys. Rev. B 29, 1558 (1984)

28. J. Friedel, private communication

29. A. Bohr, B. Mottelson, D. Pines, Phys. Rev. 110, 936 (1958); S. Belyaev, Mat. Phys. Medd. Dan. Selsk. 31, 131 (1959); A.Migdal, Nucl. Phys. 13, 655 (1959)

30. P. Ring, P. Schuck, *The Nuclear Many-Body Problem*, Springer, New York (1980)

31. J. Labbe, S.Barisic, and J. Friedel, Phys.Rev.Lett. 19, 1039 (1967)

32. H. Parmenter, Phys. Rev. 166, 392 (1968)

33. V. Kresin and B. Tavger, Sov. Phys. – JETP 23, 1124 (1966)

34. C. Owen and D. Scalapino, Physica 55, 691 (1971)

35. V.Z. Kresin, H. Gutfreund, W.Little, Solid State Commun. 51, 339 (1984)

36. V.Z. Kresin, Phys. Lett. A 122, 434 (1987)

37. A. Abrikosov, L. Gor'kov, I. Dzyaloshinski, *Methods of Quantum Field Theory in Statistical Physics* (Dover, New York, 1975)

38. L. Gor'kov, Sov. Phys. – JETP 7, 505 (1958)

39. D. Scalapino, in *Superconductivity*, ed. by R.Parks, p.449 (Dekker, New York, 1969)

40. G.Grimvall, *The Electron-Phonon Interaction in Metals*, North-Holland, Amsterdam (1981)

41. G. Eliashberg, JETP <u>12</u>,1000 (1961).

42. A.B. Migdal, *Theory of Finite Fermi Systems and Application to Atomic Nuclei* (Wiley, New York, 1967)

43. C. Yannouleas and R. A. Broglia, Annals of Physics <u>217</u>, 105 (1992)

44. M.Cohen, in *Superconductivity*, R.Parks, Ed., p.615 (Dekker, New York, 1969)

45. W.McMillan, Phys. Rev. <u>167</u>, 331 (1968)

46. A. Migdal, *Qualitative Methods in Quantum Theory*, Pergamon Press, New York (2000)

47. L. Landau and E. Lifshitz, *Quantum Mechanics*, Ch.VI, Pergamon, New York (1988)

48. V.V. Kresin and Yu. N. Ovchinnikov, Phys. Rev. B <u>73</u>, 115412 (2006).

49. M. Tinkham, *Introduction to Superconductivity*, McGraw-Hill, New York (1996)

50. A. Larkin and A. Varlamov, *Theory of Fluctuations in Superconductors*, Oxford Univ. Press, New York (2004)

51. F. Catara, P. Chomaz, N.Van Giai, Z. Phys.D <u>33</u>,219(1995)

52. E. Wolf, *Principles of Electron Tunneling Spectroscopy* (Oxford, New York, 1985)

53. G. Simmons and H. Wang*, Single Crystal Elastic Constants and Calculated Aggregate Properties*, MIT Press, Cambridge (1971)

54. T.Dietrich, T. Doppner, J. Braune, J. Tiggesbaumker, K.-H. Meiwes-Broer, Phys. Rev. Lett. <u>86</u>, 4807 (2001)

55. G. Wriggle, M. Astruk Hoffman, B. von Issendorff, Phys. Rev. A $\underline{65}$, 063201 (2002); Eur. Phys. J. D $\underline{24}$, 23 (2003)

56. B. von Issendorff and O. Cheshnovsky, Annu. Rev. Phys. Chem. $\underline{56}$, 549 (2005)

57. We are grateful to O. Cheshnovsky for bringing our attention to this point.

58. L. Adams, B. Lang, A. Goldman, Phys. Rev. Lett. $\underline{95}$, 146804 (2005)

59. J. Hagel, M. Kelemen, G. Fisher, B. Pilawa, J. Wosnitza, E. Dormann, H. v. Lohneysen, A. Schnepf, H. Schnockel, U. Neisel, J. Beck, J. Low Temp. Phys. $\underline{129,}$ 133 (2002)

60. O. Bakharev, D. Bono, H. Brom, A. Schnepf, H. Schnockel, L. de Longh, Eur. Phys. J. D $\underline{24}$, 101 (2003); cond/mat 0511612

61. A. Schindler, R. Konig, T. Herrmannsdorfer, H. Braun, G.Eska, D. Gunther, M. Meissner, R. Wahl, W. Pompe, Europhys. Lett. $\underline{58}$, 885 (2002)

62. T. Bergmann and T. Martin, J. Chem. Phys. $\underline{90}$, 2848 (1989)

**Figure Caption**

**Fig.1**. a) Electronic energy spectrum in absence of pairing for a "magic" cluster with N=198 calculated in the jellium model [27]. The shells are seen to be not equidistant; b) The single-electron spectrum for a slightly deformed clusters with an unfilled $h$ shell (e.g., N=166). The levels are classified by the projection of angular momentum. As shown in the text, pairing can drastically alter the spectrum, since the gap parameter can be on the order of, or even exceed, the single-electron level spacing near $E_F$.

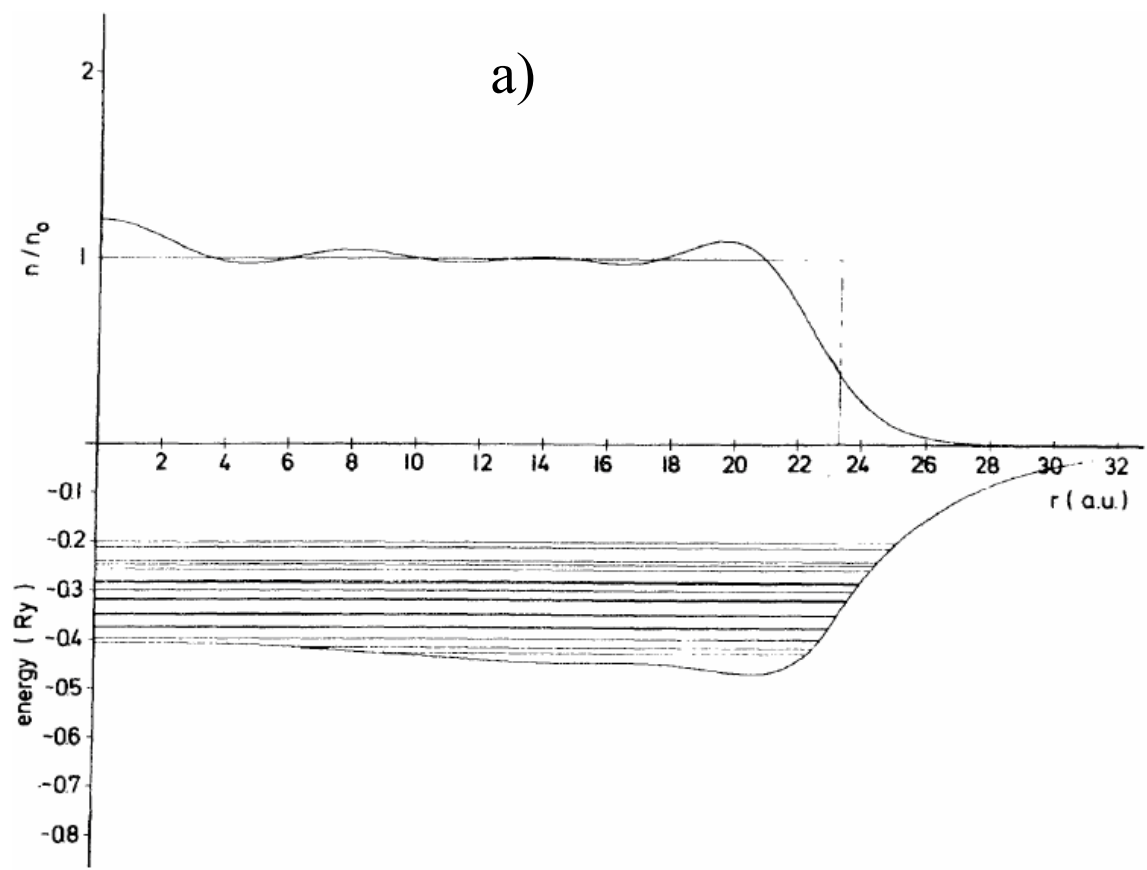
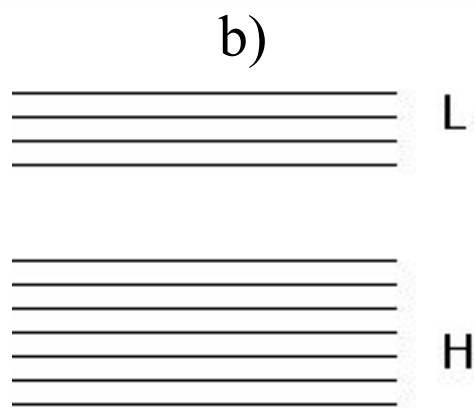

Fig. 1